\documentclass[twocolumn,aps,prl,showpacs,superscriptaddress,groupedaddress]{revtex4}
\usepackage{tikz}
\usetikzlibrary{arrows}
\usepackage{tikz-timing}
\usetikzlibrary{arrows.meta,decorations.pathmorphing,decorations.pathreplacing,positioning,shapes,fadings}
\usepackage{graphicx}
\usepackage{braket}
\usepackage{hyperref}
\usepackage{circuitikz}
\graphicspath{{pics/}}
\usepackage{amsmath}

\usepackage[colorinlistoftodos]{todonotes}
\newsavebox{\mycircuita}
\sbox{\mycircuita}{%
        \begin{circuitikz}[scale=0.375,font={\fontfamily{phv}\selectfont\footnotesize},
        	capacitor/.append style={bipoles/length=0.7cm,bipoles/thickness=1},
            inductor/.append style={bipoles/length=0.7cm,bipoles/thickness=1},
            resistor/.append style={bipoles/length=0.6cm,bipoles/thickness=1},
            transmission line/.append style={bipoles/length=.7cm,bipoles/thickness=1},
            squid/.append style={bipoles/length=.7cm,bipoles/thickness=1}
            ]
            \filldraw[fill=orange!20!white, draw=orange!20!white] (2.25,-0.5) rectangle (-2.25,3.25);
            \filldraw[fill=orange!20!white, draw=orange!20!white] (3.75,-2.25) rectangle (2.25,-8.75);     
            \filldraw[fill=cyan!20!white, draw=cyan!20!white] (3.75,-1.8) rectangle (-0.5,-9.3);   
                  \filldraw[fill=red!20!white, draw=red!20!white] (-5.5,-4.5) rectangle (-2.0,-6.5);   
        	\draw
  			(5,0) to[capacitor,/tikz/circuitikz/bipoles/length=0.5cm-, -,l=$C_{in}$] (2.0,0)
            to [-] (0.0,0.0) to[-] (-1,0) to[-] (-2.0,0.0)  to[capacitor,/tikz/circuitikz/bipoles/length=0.8cm-, -,l_=$C_{out}$]  (-5.0,0)
            (0.0,0.0) to [capacitor, -,l_=$C_{C}$](0,-2.0) 
            (-1.5,0.0) to [inductor, -,l_=$L_{R}$] (-1.5,3.0)
            (1.5,0.0) to [capacitor, -,l_=$C_{R}$] (1.5,3.0) to [-] (-1.5,3.0)
            (0.0,3.0) to node[ground,scale=0.3]{} (0.0,2.5)
            (0.0,-9.0) node[ground,scale=0.3]{} (0,-10.0)
            (-3,-2) to [-] (3,-2)
            (-3,-2) to [capacitor, -,l_=$C_{q}$] (-3,-9)
            
            (0.0,-2) to [barrier,/tikz/circuitikz/bipoles/length=0.7cm-,l_=$E_{J}$] (0.0,-9)
            (3.0,-2) to [barrier,/tikz/circuitikz/bipoles/length=1.0cm-] (3.0,-3.5)
            (3.0,-3.0) to [barrier,/tikz/circuitikz/bipoles/length=1.0cm-] (3.0,-4.0)
            (3.0,-4.0) to [barrier,/tikz/circuitikz/bipoles/length=1.0cm,-] (3.0,-4.8)
            
            (3.0,-6.8) to [barrier,/tikz/circuitikz/bipoles/length=1.0cm-,-] (3.0,-7.0)
            (3.0,-7.0) to [barrier,/tikz/circuitikz/bipoles/length=1.0cm-,-] (3.0,-8.0)
            (3.0,-8.0) to [barrier,/tikz/circuitikz/bipoles/length=1.0cm-,-] (3.0,-8.5)
            (3.0,-9.0) to [-] (-3,-9.0)
            (2.5,-5.77) node[anchor=north,rotate=90] {$\cdots$};
  \node[scale=1] at (1.5,-5.5) {$\Phi_{\mathrm{ext}}$};
    \node[scale=1] at (4.5,-5.5) {$L_{jA}$};
        \end{circuitikz}
    }

\newsavebox{\mycircuitb}
\sbox{\mycircuitb}{%
        \begin{circuitikz}[scale=0.35,font={\fontfamily{phv}\selectfont\footnotesize},
        	capacitor/.append style={bipoles/length=0.7cm,bipoles/thickness=1},
            inductor/.append style={bipoles/length=0.7cm,bipoles/thickness=1},
            resistor/.append style={bipoles/length=0.6cm,bipoles/thickness=1},
            transmission line/.append style={bipoles/length=.7cm,bipoles/thickness=1},
            squid/.append style={bipoles/length=.7cm,bipoles/thickness=1}
            ]
            
        	\draw

            (0.0,0.0) to [capacitor, -,l_=$C_{C}$](0,-2.0) 
            (0,0) to [-] (-8.0,0.0)
            (-8.0,0.0) to [capacitor, -,l_=$C_{R}$](-8.0,-5.0)
            (-5.5,0.0) to [inductor, -,l_=$L_{R}$] (-5.5,-5.0)
            (-8.0,-5.0) to [-] (-3.0,-5.0)
            (-5.5,-5.0) node[ground,scale=0.3]{} (-5.5,-5.5)
            
            (-3,-2) to [-] (3,-2)
            (-3,-2) to [capacitor, -,l_=$C_{1}$] (-3,-4.5) 
            to [capacitor, -,l_=$C_{2}$] (-3,-9)
            
            (0.0,-2) to [barrier,/tikz/circuitikz/bipoles/length=0.7cm-,l_=$E_{J}$] (0.0,-9)
            (3.0,-2) to [barrier,/tikz/circuitikz/bipoles/length=1.0cm-] (3.0,-4.0)
            (3.0,-2.6) to [-] (4.0,-2.6)
            (4.0,-2.6) to [capacitor,-] (4.0,-3.4)
            (3.0,-3.4) to [-] (4.0,-3.4)
            (3.0,-3.8) to [-] (5.0,-3.8)
            (5.0,-3.8) to [capacitor,/tikz/circuitikz/bipoles/length=0.4cm-] (6.0,-3.8)
            (6.0,-3.8) node[ground,scale=0.3]{} (6.0,-4.0)
            (3.0,-4.0) to [barrier,/tikz/circuitikz/bipoles/length=1.0cm,-] (3.0,-4.8)
            (3.0,-4.0) to [-] (4.0,-4.0)
            (4.0,-4.0) to [capacitor,-] (4.0,-4.8)
            (3.0,-4.8) to [-] (4.0,-4.8)
            (3.0,-5.2) to [-] (5.0,-5.2)
            (5.0,-5.2) to [capacitor,/tikz/circuitikz/bipoles/length=0.4cm-] (6.0,-5.2)
            (6.0,-5.2) node[ground,scale=0.3]{} (6.0,-5.0)
            
            (3.0,-6.8) to [barrier,/tikz/circuitikz/bipoles/length=1.0cm-,-] (3.0,-7.0)
            (3.0,-6.5) to [-] (4.0,-6.5)
            (4.0,-6.5) to [capacitor,-] (4.0,-7.3)
            (3.0,-7.3) to [-] (4.0,-7.3)
            (3.0,-7.6) to [-] (5.0,-7.6)
            (5.0,-7.6) to [capacitor,/tikz/circuitikz/bipoles/length=0.4cm-] (6.0,-7.6)
            (6.0,-7.6) node[ground,scale=0.3]{} (6.0,-7.8)

            (3.0,-8.0) to [barrier,/tikz/circuitikz/bipoles/length=1.0cm-,-] (3.0,-8.5)
            (3.0,-7.8) to [-] (4.0,-7.8)
            (4.0,-7.8) to [capacitor,-] (4.0,-8.6)
            (3.0,-8.6) to [-] (4.0,-8.6)
            (3.0,-8.8) to [-] (5.0,-8.8)
            (5.0,-8.8) to [capacitor,/tikz/circuitikz/bipoles/length=0.4cm-] (6.0,-8.8)
            (6.0,-8.8) node[ground,scale=0.3]{} (6.0,-9.0)
            (3.0,-9.0) to [-] (-3,-9.0) 
            (5.8,-3.5) to [-] (6.6,-3.2) 
            (4.2,-2.6) to [-] (5.0,-1.9) 
            (2.5,-2.6) to [-] (2.0,-1.4) 
            (2.5,-5.77) node[anchor=north,rotate=90] {$\cdots$};
  \node[scale=1] at (1.5,-5.5) {$\Phi_{\mathrm{ext}}$};
  \node[scale=1] at (7.0,-3.0) {$C_{g_i}$};
  \node[scale=1] at (5.0,-1.5) {$C_{jA_i}$};
  \node[scale=1] at (2.0,-1.0) {$L_{jA_i}$};
  
          \node[scale=1] at (-5.5,0.5) {$\varphi_{R}$};
       \node[scale=1] at (1.0,-2.75) {$\varphi_{1}$};
       \node[draw,circle,fill=black,scale=0.2] at (0.0,-2.0) {};
\node[scale=1] at (0.0,-9.75) {$\varphi_{1}-\varphi_{Q}$};
       \node[draw,circle,fill=black,scale=0.2] at (0.0,-9.0) {};
             \node[draw,circle,fill=black,scale=0.2] at (-5.5,0.0) {};
        \end{circuitikz}
    }
\newsavebox{\mycircuitc}
\sbox{\mycircuitc}{%
        \begin{circuitikz}[scale=0.35,font={\fontfamily{phv}\selectfont\footnotesize},
        	capacitor/.append style={bipoles/length=0.7cm,bipoles/thickness=1},
            inductor/.append style={bipoles/length=0.7cm,bipoles/thickness=1},
            resistor/.append style={bipoles/length=0.6cm,bipoles/thickness=1},
            transmission line/.append style={bipoles/length=.7cm,bipoles/thickness=1},
            squid/.append style={bipoles/length=.7cm,bipoles/thickness=1}
            ]
            
        	\draw
            (-3,0) to [capacitor, -,l_=$\tilde{C}_{C}$] (-7.0,0.0)
                        (-7,0) to [-] (-10.0,0.0)
            (-10.0,0.0) to [capacitor, -,l_=$\tilde{C}_{R}$](-10.0,-7.0)
            (-7.0,0.0) to [inductor, -,l_=$L_{R}$] (-7.0,-7.0)
            (-10.0,-7.0) to [-] (-3.0,-7.0)
            (-3,-7.0) node[ground,scale=0.3]{} (-3,-9.5)     
            (-3,0) to [-] (3,0)
            (-3,0) to [capacitor, -,l_=$\tilde{C}_{Q}$] (-3,-7)
            (0.0,0) to [barrier,/tikz/circuitikz/bipoles/length=0.7cm-,l_=$E_{J}$] (0.0,-7)
            (3.0,0) to [inductor, -,] (3.0,-7.0)
            (3.0,-7.0) to [-] (-3,-7.0);
  \node[scale=1] at (1.5,-3.5) {$\Phi_{\mathrm{ext}}$};
    \node[scale=1] at (4.5,-3.5) {$L_{jA}$};
      \node[scale=1] at (-7.0,0.5) {$\varphi_{R}$};
       \node[scale=1] at (-0.0,0.5) {$\varphi_{Q}$};
              \node[draw,circle,fill=black,scale=0.2] at (0.0,0.0) {};
 \node[draw,circle,fill=black,scale=0.2] at (-7.0,0.0) {};
        \end{circuitikz}
    }

\begin{document}
\begin{abstract}

We realize a $\Lambda$ system in a superconducting circuit, with metastable states exhibiting lifetimes up to 8\,ms.  We exponentially suppress the tunneling matrix elements involved in spontaneous energy relaxation by creating a ``heavy'' fluxonium, realized by adding a capacitive shunt to the original circuit design.  The device allows for both cavity-assisted and direct fluorescent readout, as well as state preparation schemes akin to optical pumping.  Since direct transitions between the metastable states are strongly suppressed, we utilize Raman transitions for coherent manipulation of the states.

\end{abstract}

\title{Realization of a $\Lambda$ system with metastable states of a capacitively-shunted fluxonium}

\author{N. Earnest}
\email{nearnest@uchicago.edu}
\affiliation{The James Franck Institute and Department of Physics, University of Chicago, Chicago, Illinois 60637, USA}
\author{S. Chakram}

\affiliation{The James Franck Institute and Department of Physics, University of Chicago, Chicago, Illinois 60637, USA}
\author{Y. Lu}
\affiliation{The James Franck Institute and Department of Physics, University of Chicago, Chicago, Illinois 60637, USA}
\author{N. Irons}
\affiliation{Department of Physics and Astronomy, Northwestern University, Evanston, Illinois 60208, USA}
\author{R. K. Naik}
\affiliation{The James Franck Institute and Department of Physics, University of Chicago, Chicago, Illinois 60637, USA}
\author{N. Leung}
\affiliation{The James Franck Institute and Department of Physics, University of Chicago, Chicago, Illinois 60637, USA}
\author{L. Ocola} \affiliation{Argonne National Laboratories, Center for Nanoscale Materials, Argonne, Illinois 60439, USA}

\author{D.A. Czaplewski} \affiliation{Argonne National Laboratories, Center for Nanoscale Materials, Argonne, Illinois 60439, USA}
\author{B. Baker}
\affiliation{Department of Physics and Astronomy, Northwestern University, Evanston, Illinois 60208, USA}
\author{Jay Lawrence}
\affiliation{Department of Physics, Dartmouth College, Hanover, New Hampshire 03755, USA}
\author{Jens Koch}
\affiliation{Department of Physics and Astronomy, Northwestern University, Evanston, Illinois 60208, USA}
\author{D. I. Schuster}
\email{david.schuster@uchicago.edu}
\affiliation{The James Franck Institute and Department of Physics, University of Chicago, Chicago, Illinois 60637, USA}
\maketitle
\thispagestyle{empty}

\tikzfading[name=fade out,inner color=transparent!0,outer color=transparent!100]

Quantum computation with superconducting circuits has seen rapid progress over the past decade~\cite{kelly2015state,kandala2017hardware,ofek2016extending} largely due to improvements in qubit coherence~\cite{wallraff2004strong, koch2007, paik2011observation}. Performing large-scale quantum computation, error correction and simulation, will require significantly longer coherence times~\cite{fowler2012surface,devoret2013superconducting}. Fermi's golden rule states that qubit lifetimes are governed by two factors: (1) the noise spectral density associated with environmental degrees of freedom; (2) the transition matrix elements, which are determined by the qubit wavefunctions. To date, improvements in superconducting qubit lifetimes have primarily been achieved by modifying the noise spectral density -- for example, by filtering~\cite{martinis1987experimental, reed2010fast} and by minimizing the contributions of two-level systems~\cite{paik2011observation, martinis2014ucsb}. We demonstrate a complimentary approach, making the qubit insensitive to environmental noise by reducing the transition matrix elements. This leads to qubit lifetimes as high as 8 ms, and realizes a $\Lambda$ system analogous to those commonly found in atomic systems.

Most current superconducting qubit architectures are based on variants of the transmon qubit~\cite{koch2007, takita2016demonstration,barends2014superconducting,versluis2016scalable}. Transmons have large dipole matrix elements, simple selection rules, and a small non-linearity, sufficient to resolve the lowest energy levels as the qubit states. In contrast, flux qubits~\cite{mooij1999josephson,chiorescu2004coherent,zhou2004suppression} have a large nonlinearity, rich level structure, and selection rules that can be finely engineered to yield a smooth trade-off between decay matrix elements and gate fidelities. With the realization of a linear superinductance and the fluxonium qubit~\cite{manucharyan2009fluxonium}, this class of qubits has seen enhanced lifetimes and reduced flux-noise induced decoherence~\cite{pop2012experimental}. These properties make the fluxonium a promising system for engineering a $\Lambda$ system.

Traditionally, a $\Lambda$ system is comprised of a ground and metastable excited state, coherently coupled through a third intermediate state. $\Lambda$ systems are ubiquitous in atomic physics, realized using a combination of selection rules~\cite{nicholson2015systematic}, relative strengths of optical-dipole and microwave hyperfine matrix elements~\cite{monroe1995demonstration},  and large differences in frequency scales in conjunction with the 3D density of states (decay rate $\gamma \propto \nu^{3}$)~\cite{weisskopf1930calculation}.  Superconducting qubits are typically not protected by symmetry-based selection rules, and possess a much smaller dynamic range of frequency scales and a 1D density of states ($\gamma \propto \nu$), making it more challenging to realize the metastability required to explore the physics associated with $\Lambda$ systems. Previous work in cQED has utilized the Purcell effect~\cite{houck2008controlling} to modify the density of states and explore multi-tone coherent interactions of three-level systems~\cite{novikov2016raman,abdumalikov2010electromagnetically}. 

In this work, we present a $\Lambda$ system in a capacitively shunted fluxonium circuit: the heavy fluxonium. The added capacitance further localizes the lowest energy states, exponentially suppressing the dipole matrix elements and boosting the metastable state lifetime to 8\,ms. The suppressed matrix elements make controlled population transfer to this state a challenge, but we surmount this by using multi-tone Raman transitions in the $\Lambda$ system to perform coherent operations with substantial improvement in gate fidelities relative to direct transitions. 

\begin{figure}[h!]
    \begin{tikzpicture}
 \node[anchor=north west] (image) at (0,0) {\includegraphics[width=.5\textwidth]{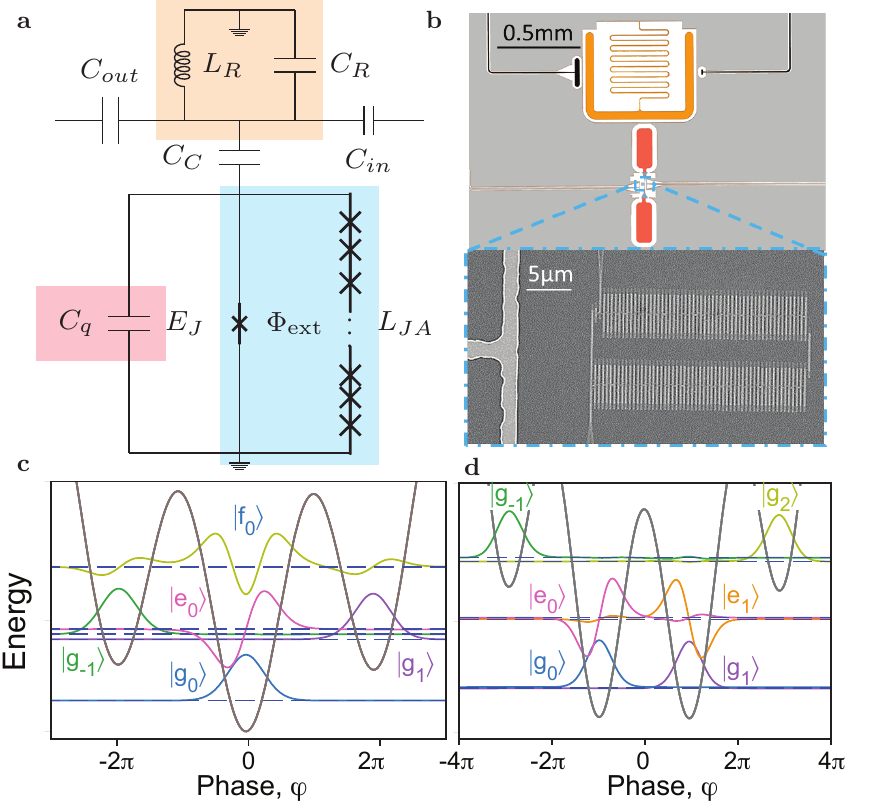}};
    \end{tikzpicture}
    \caption{a) Equivalent circuit diagram of the heavy fluxonium capacitively coupled to a readout resonator (Supplementary Information). b) Scanning Electron Microscope image of the device, with a magnified view of the 100 Josephson junction array, fabricated using the bridgeless method detailed in ~\cite{lecocq2011junction}. c) Simulated Potential energy landscape/wavefunctions at $\Phi_{{\rm ext}} = 0.02\, \Phi_0$ demonstrating localized wavefunctions in three wells. d) Simulated Potential energy landscape at $\Phi_{{\rm ext}} = 0.51 \Phi_0$, where $\ket{g_0}$ and $\ket{g_1}$ are nearly degenerate. \label{fig1}}
\end{figure}

\begin{figure}
\includegraphics[width=\columnwidth]{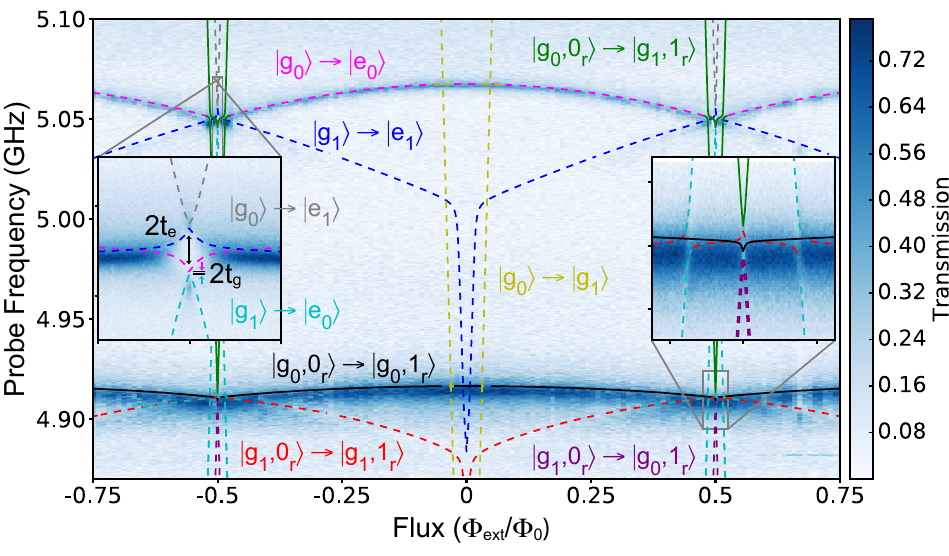}
\caption{Single-tone spectroscopy of the fluxonium-resonator system in the vicinity of the resonator and primary plasmon transition frequencies.  Dashed lines indicate simulated energy levels of the coupled system based on device parameters extracted from fits to single and two-tone spectra. Transitions that change rapidly with flux are inter-well fluxon transitions, while the flatter transitions are intra-well plasmon transitions.  Left inset: interference due to coupling between the ground and excited states of neighbouring wells.  Right inset: features associated with fluxon transitions crossing with the resonator. The spectrum is normalized by the transmission amplitude of the bare resonator (Supplementary Information). 
\label{singspec}}
\end{figure}

The heavy fluxonium circuit (Fig.~\ref{fig1}a,b) consists of a small-area Josephson junction connected in parallel to a capacitance ($C_{q}$) and a large superinductor ($L_{JA}$), realized as an array of 100 large-area Josephson junctions. To ensure idealized inductive behavior of the array, the length and individual junction size must satisfy several conditions as explained in ~\cite{manucharyan2009fluxonium}. Once these conditions are satisfied, the Hamiltonian of the fluxonium is given by:
\begin{equation}
H_f = -4 E_C \frac{d^2}{d \varphi^{2}} - E_J \cos (\varphi - 2\pi \Phi_{{\rm ext}}/\Phi_0)+\frac{1}{2}E_L \varphi^2 
\label{eq:fluxHam}
\end{equation}
where $E_C = e^{2}/2C_{q}$ is the charging energy, $E_J$ the Josephson energy of the small junction, and $E_{L} = \Phi^{2}_{0}/2L_{JA}$ the inductive energy of the Josephson junction array. In contrast to earlier fluxonium devices~\cite{manucharyan2009fluxonium}, the heavy fluxonium shunts the small junction with a large capacitance (43 fF, dashed red squares in Fig.~1b). This results in a reduced $E_{C}/h = 0.46 $\,GHz, increases the effective mass of the phase degree of freedom, and produces quasi-localized states in the different wells of the potential (see Fig.~\ref{fig1}c,d). The other circuit parameters, $E_J/h = 8.11$\,GHz and $E_L/h = 0.24$\,GHz, are comparable to those in previous fluxonium devices. We label states by their fluxoid number $\left\{-1, 0, 1\right\}$ (number of flux quanta in the loop formed by the junctions), and by the plasmon levels within that well $\left\{\ket{g},\ket{e},\ket{f}\right\}$. The heavy fluxonium allows for two types of transitions: intra-well \emph{plasmons} (e.g., $\ket{g_{0}} \leftrightarrow \ket{e_{0}}$), and inter-well \emph{fluxons} (e.g., $\ket{g_{0}} \leftrightarrow \ket{g_{1}}$).

\begin{figure*}
\begin{tikzpicture}
\node[anchor=north west] (image) at (-1.5,0) {\includegraphics[width=\textwidth]{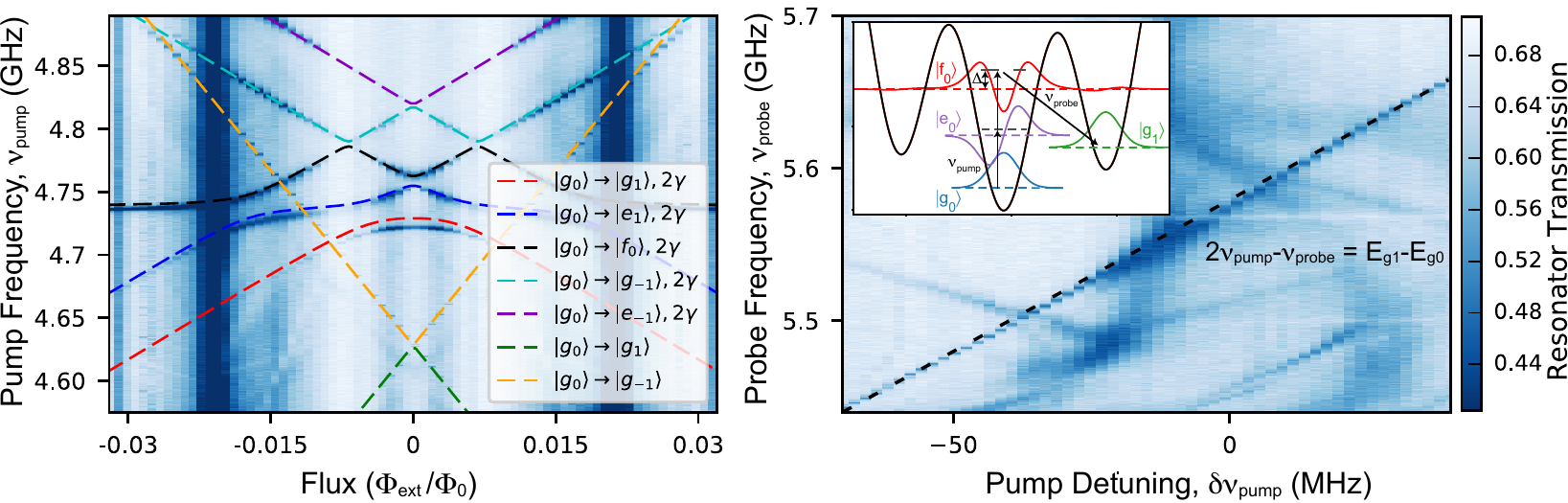}};
\node[scale=0.8] at (0.25,0) {\textbf{a}};
\node[scale=0.8] at (9.25,0) {\textbf{b}};
\end{tikzpicture}
  \caption{a) Two-tone spectroscopy showing direct fluxon transitions (orange and green lines) and two-photon transitions to the two-excitation manifold ($\ket{f_0}$, $\ket{e_{-1}}$,$\ket{e_1}$). State labeling for the transitions is valid for $\Phi_{\mathrm{ext}} > 0$. The $\ket{f_{0}}$ level serves as the intermediate state in a $\Lambda$ system comprising the ground state $\ket{g_{0}}$ and the metastable $\ket{g_{1}}$ state, and assists in Raman transitions.  b) Pump-probe spectroscopy of Raman transitions between $\ket{g_{0}}$ and  $\ket{g_{1}}$ as a function of pump (near $\ket{g_{0}} \rightarrow \ket{f_0}$-$2\gamma$ transition) and probe frequency (near $\ket{g_{1}} \rightarrow \ket{f_0}$).  The Raman transition is seen when $2\nu_{\mathrm{pump}}-\nu_{\mathrm{probe}} = E_{g_1}-E_{g_{0}}$, represented by the dashed line. The upper-left inset shows wavefunctions of the states involved in the transition. The intermediate $\ket{f_0}$ state couples to $\ket{g_0}$ via a two-photon process, and has a small amplitude in the right well, with a direct dipole-allowed transition to the metastable $\ket{g_{1}}$ state. The dashed lines are simulated energy levels of the fluxonium-resonator system. The colorbar is normalized by the transmission of the bare resonator.
\label{Lambda}}
\end{figure*}

Inter-well transitions involve states with wavefunctions such as $\psi_{g_0}(\varphi)$ and $\psi_{g_1}(\varphi)$, which are disjoint. Accordingly, matrix elements $\int d\varphi\,\psi^{*}_{g_{1}}(\varphi)\hat{O}\,\psi_{g_{0}}(\varphi)$ with respect to local operators $\hat{O}(d/d\varphi,\varphi)$ will be exponentially suppressed with $\sim \exp[-\pi^2(E_J/8E_{C})^{1/2}]$, inferred from considering the tails of displaced harmonic-oscillator wavefunctions~\cite{zhu2013asymptotic}. Consequently, the $\ket{g_{1}}$ state of the heavy fluxonium is much longer lived relative to the original fluxonium. However, the suppressed transition matrix elements also make coherent operations more challenging. This circuit resembles a recently reported design~\cite{lin2017protecting}, whose dipole moment (and thereby the fluxon transition rate) is tunable through the use of a SQUID in place of a single Josephson junction.  Unlike the fluxonium in ~\cite{lin2017protecting}, our heavy fluxonium  -- with a fixed $E_J/E_C \approx 18$ -- is sufficiently heavy to disallow coherent direct drives. We solve this issue by realizing a $\Lambda$ system between the ground state $\ket{g_0}$, the metastable state $\ket{g_1}$, and the excited state $\ket{f_0}$, and perform coherent Raman transitions between $\ket{g_0}$ and $\ket{g_1}$. 

For fast readout, the heavy fluxonium is capacitively coupled to a lossy resonator ($Q \sim 500$). The Hamiltonian of the combined system is given by~\cite{zhu2013circuit}:
\begin{equation}
H_S = H_f + h\nu_r \hat{a}^{\dagger}\hat{a} + \sum_{j,k} h g\ket{j}\bra{k} \bra{j}\hat{n}\ket{k} (\hat{a} + \hat{a}^{\dagger}),
\label{sysHam}
\end{equation}
where, $\nu_r = 4.95$\,GHz is the bare frequency of the resonator and 
$g = 76$\,MHz is the coupling between the resonator and fluxonium 
(as extracted from fits to spectra). $\hat{n}$ is the charge operator of the fluxonium and controls the transition rates arising from driving on the input port ($C_{in}$ in Fig.~\ref{fig1}a).

Single-tone spectroscopy (Fig.~\ref{singspec}) reveals both the resonator photon and the plasmon transitions. The curvature of the plasmon transitions arises from flux-induced distortion of the well (Supplementary Information), and allows one to easily distinguish between wells (blue and magenta lines in Fig.~\ref{singspec}). Furthermore, the strong hybridization of the plasmon and resonator (detuned by up to $155$\,MHz) allows for fluorescent readout of the metastable state, over the entire flux range, through cycling the plasmon transition of the metastable state many times, similar to quantum non-demolition measurements of single trapped ions and atoms~\cite{neuhauser1980localized, schlosser2001sub}.   

The tunnel splitting between the wells can be directly observed in the plasmon spectrum at $\Phi_{\rm ext} = \Phi_0/2$. At this flux location, there are two identical wells with degenerate ground and first excited states. This results in the feature shown in the left inset of Fig.~\ref{singspec}, where the interference of the levels results in a unique rhombus-shaped avoided crossing.  The separation of the level crossings forming the top and bottom corners of the rhombus (black arrows) is a direct measure of the tunnel coupling of the excited states in the well ($\ket{e_0}$ and $\ket{e_1}$), corresponding to $t_{e} \approx 7$\,MHz. The tunnel splitting between ground states is smaller than the linewidth of the plasmon and fluxon transitions, and is inferred from the fits to be $t_g \approx 0.42$\,MHz, one thousand times smaller than in previous experiments~\cite{manucharyan2009fluxonium}. Another set of avoided crossings is visible in the resonator transmission peak (right inset of Fig.~\ref{singspec}) at $\Phi_{\rm ext}\approx 0.5 \Phi_0$. The outer set of crossings arise from the $\ket{g_1} \rightarrow \ket{e_0}$ fluxon transition (also seen in the bottom of the left inset), while the inner crossings are formed by composite resonator/fluxon transitions: $\ket{g_1,0_{\mathrm{res}}} \rightarrow \ket{g_0,1_{\mathrm{res}}}$ and $\ket{g_0,0_{\mathrm{res}}} \rightarrow \ket{g_1,1_{\mathrm{res}}}$. The latter indicate that coupling between fluxon transitions is increased when a photon is present in the resonator (Supplementary Information for photon-assisted fluxon transitions). The fluxon transitions are (to first order) linear in flux, with slopes given by
$\partial f / \partial \Phi_{\rm ext}  \approx \pm 4 \pi^2 E_L/\Phi_0 = \pm 9.59$\,GHz/$\Phi_0$. 

Fluxon transitions, not seen in single-tone spectroscopy, can be identified via two-tone spectroscopy in which we monitor the transmission of the readout resonator while sweeping the frequency of a second drive tone.  The lines of largest slope are the single-photon inter-well fluxon transitions $\ket{g_{0}} \rightarrow \ket{g_{\pm 1}}$. The rest of the lines are two-photon transitions to the second-excited manifold of the fluxonium-resonator system, with flat features corresponding to plasmons and sloped resonances corresponding to fluxons.  Of particular importance are the two-photon features located at $\sim 4.73$\,GHz corresponding to the $\ket{g_0}\rightarrow \ket{f_{0}}$ two-photon transition, which will assist in performing coherent operations on the qubit.

The heavy fluxonium energy-level structure allows for a variety of state-preparation schemes. We can perform $T_1$ measurements from the highest fluxon transition frequency of $4.65$\,GHz down to about $3$\,GHz by directly driving the fluxon transition at high powers to realize a classically mixed state (100\,$\mu$s pulse duration).  Below this point,  we perform $T_1$ measurements using a process that is similar to optical pumping~\cite{cohen1966optical}. Through continuous cycling of the bright $\ket{g_0} \rightarrow \ket{e_0}$ plasmon in Fig.~\ref{singspec}, we take advantage of a small probability of decaying from $\ket{e_0}$ to $\ket{g_1}$ arising from the finite matrix element between these states and incoherently ``pump'' the system into the $\ket{g_{1}}$ state, and perform a typical $T_1$ measurement.  
\begin{figure}
\begin{tikzpicture}
\node[anchor=north west] (image) at (-1.5,0){\includegraphics[scale=0.35]{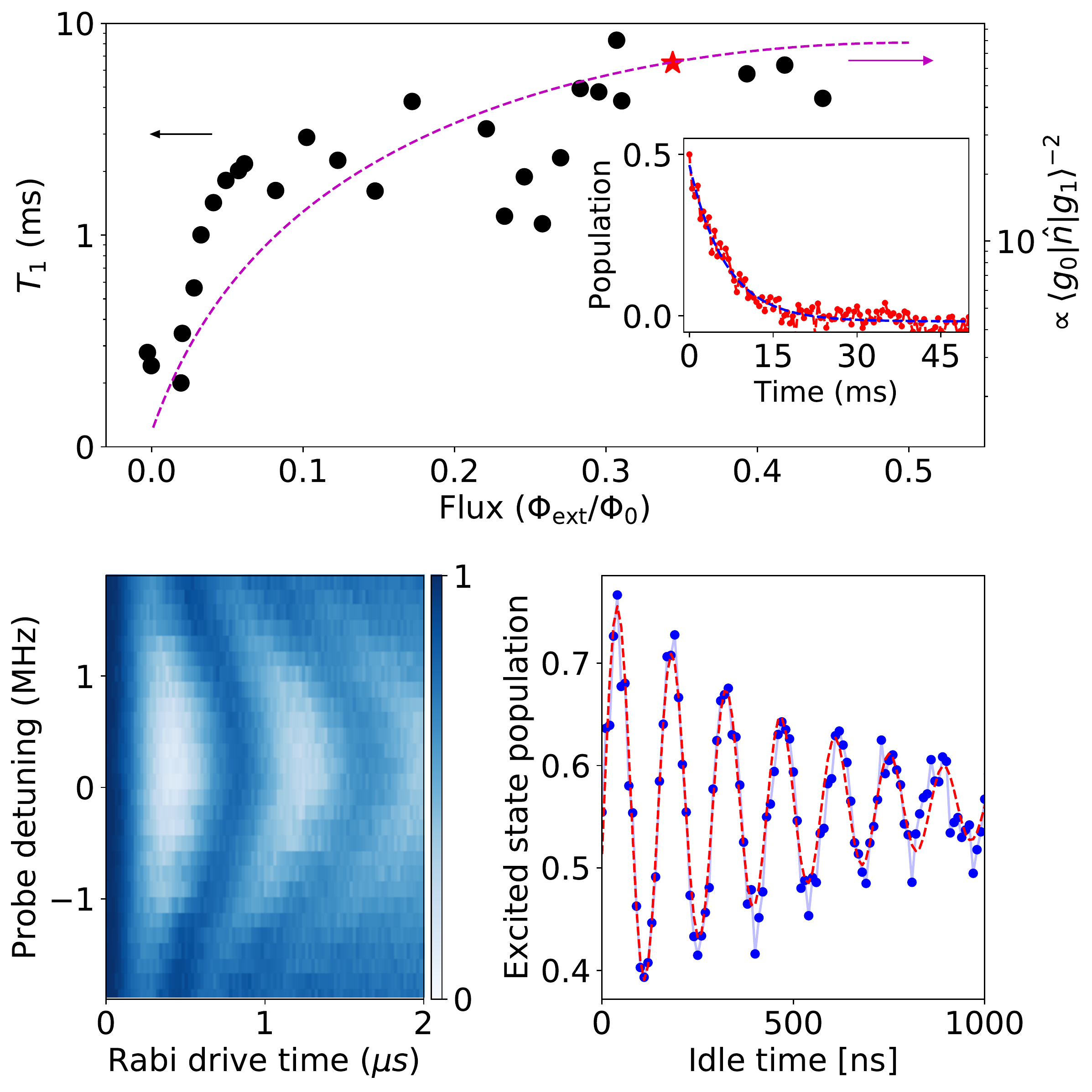}};
\node[scale=0.8] at (-0.25, 0) {\textbf{a}};
\node[scale=0.8] at (-0.25,-4.4) {\textbf{b}};
\node[scale=0.8] at (3.65,-4.4) {\textbf{c}};
\end{tikzpicture}
\caption{a) Energy relaxation time ($T_{1}$) as a function of magnetic flux measured through a combination of direct-drive to a mixed state, plasmon pumping, and Raman transitions. The dashed purple line indicates the inverse square of the charge matrix element of the fluxon transition of interest. The inset shows the $T_{1}$ decay curve of the point indicated by the red star, after driving to a mixed state following a long Raman drive. (b)
Rabi chevron obtained by detuning the probe-drive tone away from the Raman transition described in Fig.~\ref{Lambda}b  at $\Phi_{\mathrm{ext}} = 0.078\,\Phi_{0}$. The Raman transition is chosen to be 60\,MHz from the $\ket{f0}$ level and the peak $\pi$ pulse fidelity is $\sim 90 \%$.
(c) Ramsey experiment at $\Phi_{\mathrm{ext}} = 0.078\,\Phi_{0}$, obtained using $\pi/2$ pulses extracted from Rabi drive of the Raman transition resulting in a $T^{*}_{2}$ of 500-550 ns.
\label{coherence}}

\end{figure}
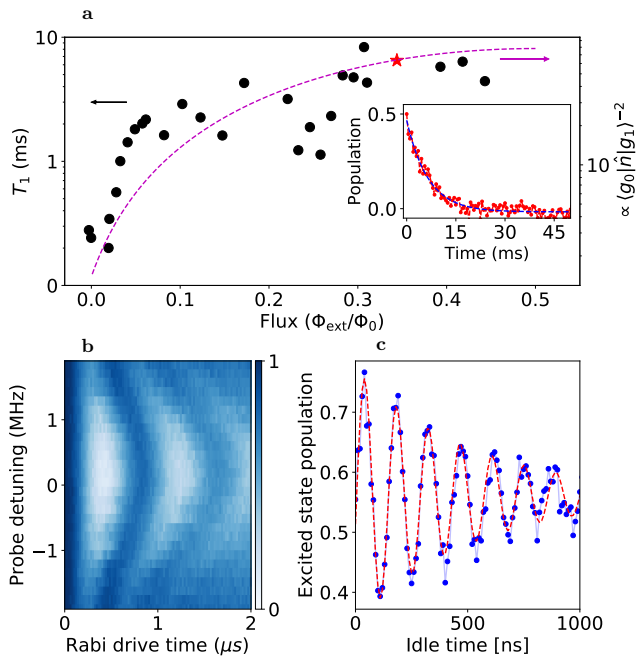

Since direct fluxon transitions are forbidden, we realize faster gates by means of Raman transitions that utilize the excited levels of the fluxonium, in analogy with atomic physics. Recently, such multi-tone transitions have been used in superconducting qubits in the context of stabilization, and coherent population trapping~\cite{chiorescu2004coherent,Arimondo1996VSpectroscopy,lu2017universal,kelly2010direct,novikov2016raman,abdumalikov2010electromagnetically}. As tunneling is suppressed exponentially by the depth of the well, it is advantageous to use higher plasmon excited states.  Of particular importance is the $\ket{g_0} \rightarrow \ket{f_0}$ transition shown in the inset of Fig.~\ref{Lambda}. Though the direct transition is disallowed by the symmetry of the wavefunctions, we can access it through a two-photon process mediated by the $\ket{e_{0}}$ level.  Further, from the inset in Fig. \ref{Lambda}b we can see that the $\ket{f_{0}}$ wavefunction has a noticeable amplitude in the right well, and $\ket{g_{1}}\rightarrow \ket{f_0}$ is dipole allowed. This indicates that we can use the $\ket{g_0}$, $\ket{f_0}$, and $\ket{g_{1}}$ states to form a $\Lambda$ system. We explore Raman transitions in this  $\Lambda$ system by sweeping the pump and probe tone frequencies in the vicinity of these transitions, as shown in Fig. \ref{Lambda}b. We find a shift in the resonator transmission when $2\nu_{\mathrm{pump}} -\nu_{\mathrm{probe}} =  E_{g_1}-E_{g_{0}}$, corresponding to the intended transfer of population from $\ket{g_{0}} \rightarrow \ket{g_{1}}$.  The Raman transition rate is related to the Rabi rates of the two Raman tones, $\Omega_{\mathrm{probe}}$ from $\ket{g_{1}}\rightarrow\ket{f_0}$, and $\Omega_{\mathrm{pump}}$ from $\ket{g_0}\rightarrow\ket{f_0}$ according to:
\begin{equation}
\Omega_{g_0 g_{1}} = \frac{\Omega_{\mathrm{probe}}\Omega^{2}_{\mathrm{pump}}}{\Delta\delta_{2\gamma}} ,
\label{RamanRate}
\end{equation}
where $\Delta = 2\nu_{\mathrm{pump}}-E_{f_0} = \nu_{\mathrm{probe}}-(E_{f_{0}} - E_{g_{1}} )$ is the detuning of the pump and probe tone from the two-photon resonance, while $\delta_{2\gamma} = E_{e_{0}} -\nu_{\mathrm{pump}}$ is the detuning of the two-photon $\ket{g_{0}}\rightarrow\ket{f_0}$ pump tone from the intermediate $\ket{e_{0}}$ state.  $\Omega_{\mathrm{pump}}$ and $\Omega_{\mathrm{probe}}$ are set by the strength of the drive and by the charge matrix elements of the $\ket{g_{1}}\rightarrow\ket{f_0}$ and $\ket{g_0}\rightarrow\ket{e_0}$ transitions, respectively. 

Having established the $\Lambda$ system and the necessary tones required to perform a Raman transition between the otherwise forbidden metastable states, we induce Rabi oscillations by simultaneously switching on resonant pump and probe drives. The pump is detuned $30$\,MHz from the two-photon $\ket{f_{0}}$ transition, and the probe frequency is chosen to be $\nu_{\mathrm{probe}} = 2\nu_{\mathrm{pump}}-\Delta E_{g_1 g_0}$. At a flux value of $0.078\, \Phi_0$ we achieve a $\pi$ pulse rate of $t_{\pi}\sim 400$\,ns with 90$\%$ contrast (Fig.~\ref{coherence}b).
While this fidelity can be further optimized in future devices, it demonstrates several orders of magnitude improvement from the direct drive which takes 100\,$\mu$s to generate a classically mixed state (Supplementary Information).  The upper limit of the Raman transition rate arises from off-resonant excitation of the resonator through the two-photon pump drive, which drives the $\ket{g_{0},1_{res}}\rightarrow \ket{g_{1},1_{res}}$ transition. 

Using these different methods, direct driving, plasmon pumping, and a three-photon Raman transition, we measure the $T_{1}$ of the device over the flux range $0 \leq |\Phi_{\mathrm{ext}}| < 0.45 \,\Phi_{0}$, as shown in Fig.~\ref{coherence}a.
Plotting the $T_{1}$ versus flux shows improvement as we move toward $0.5\,\Phi_{0}$ and follows the (inverse square of the) charge matrix elements. This indicates that the $T_1$ is limited by a charge-based loss mechanism, as was also observed in the recent work on a similar fluxonium device~\cite{lin2017protecting}. Furthermore, we successfully measure the coherence of the fluxon transition using a standard Ramsey sequence (Fig. \ref{coherence}c) with $\frac{\pi}{2}$ pulses extracted from Rabi oscillations (Fig. \ref{coherence}b).  The $T^{*}_{2}$ is measured to be $500-550$\,ns. Using the measured flux slope and assuming a $1/f$ form, this corresponds to a flux noise spectral density $S_{\phi}(1\,\mathrm{Hz})=1.3\,\mathrm{\mu\Phi_{0}/\sqrt{Hz}}$, comparable to the flux noise measured for tunable transmons with similar magnetic shielding. This indicates that $T^{*}_{2}$ should be improved by increasing the chain inductance, since the transition flux slope is given by $\partial f / \partial \Phi_{\rm ext} \sim 1/L$. A spin-echo experiment using Raman transition-based $\pi/2$ and $\pi$ pulses gives a $T_{2,\mathrm{echo}}$ of 1.3$\,\mu$s with a single inserted $\pi$ pulse.

In summary, we have realized a heavy fluxonium in a 2D cQED architecture, with  metastable states exhibiting lifetimes of several milliseconds over a broad range of flux values, likely limited by a charge-based loss mechanism. We study the coherence of the device by state preparation schemes that use the rich energy level structure of the device, including a process analogous to optical pumping. We perform coherent operations on the long-lived metastable states using a three-photon Raman transition using an excited plasmon level as the intermediate state, realizing single-fluxon gates  ($t_{\pi} \sim 400$\,ns)  that are several orders of magnitude faster than directly driving the fluxon transition with comparable microwave powers. Additionally, the relative proximity of the plasmon and readout resonator allow for photon and plasmon-mediated transitions, that could be useful for high-fidelity fluorescent readout and photon detection schemes with cQED (Supplementary Information).  

In future work, we seek to improve the speed and fidelity of inter-well transitions, by increasing the lifetime of the plasmon states, by using more sophisticated multi-tone techniques\cite{zhou2017accelerated}, and by increasing the inductance to reduce dephasing rates. The fabrication techniques developed here will be useful for other types of protected qubits including the $0-\pi$~\cite{dempster2014understanding} and Josephson rhombus chain qubits~\cite{bell2014protected}.

\begin{acknowledgements}
We thank R. Ma for help with the pulsed measurement setup. This material is based upon work supported by the Army Research Office under (W911NF-15-1-0421), and by the National Science Foundation Graduate Research Fellowship under Grant No. DGE-1144082.  This work was partially supported by the University of Chicago Materials Research Science and Engineering Center, which is funded by the National Science Foundation under award number DMR-1420709. Use of the Center for Nanoscale Materials, an Office of Science user facility, was supported by the U. S. Department of Energy, Office of Science, Office of Basic Energy Sciences, under Contract No. DE-AC02-06CH11357. We gratefully acknowledge support from the David and Lucille Packard Foundation.
\end{acknowledgements}

\bibliography{sample.bib}

\title{Realization of a $\Lambda$ system with metastable states of a capacitively-shunted fluxonium: Supplementary Information}

\maketitle
\thispagestyle{empty}

\tikzfading[name=fade out,inner color=transparent!0,outer color=transparent!100]

\maketitle
\appendix 
\renewcommand{\thefigure}{S\arabic{figure}}
\setcounter{figure}{0}
\renewcommand\theequation{S\arabic{equation}}
\setcounter{equation}{0}

\section{Device design and fabrication}
\begin{figure}
\begin{tikzpicture}
\node[anchor=north west] (image) at (-1.5,0) {\includegraphics[width=0.45\textwidth]{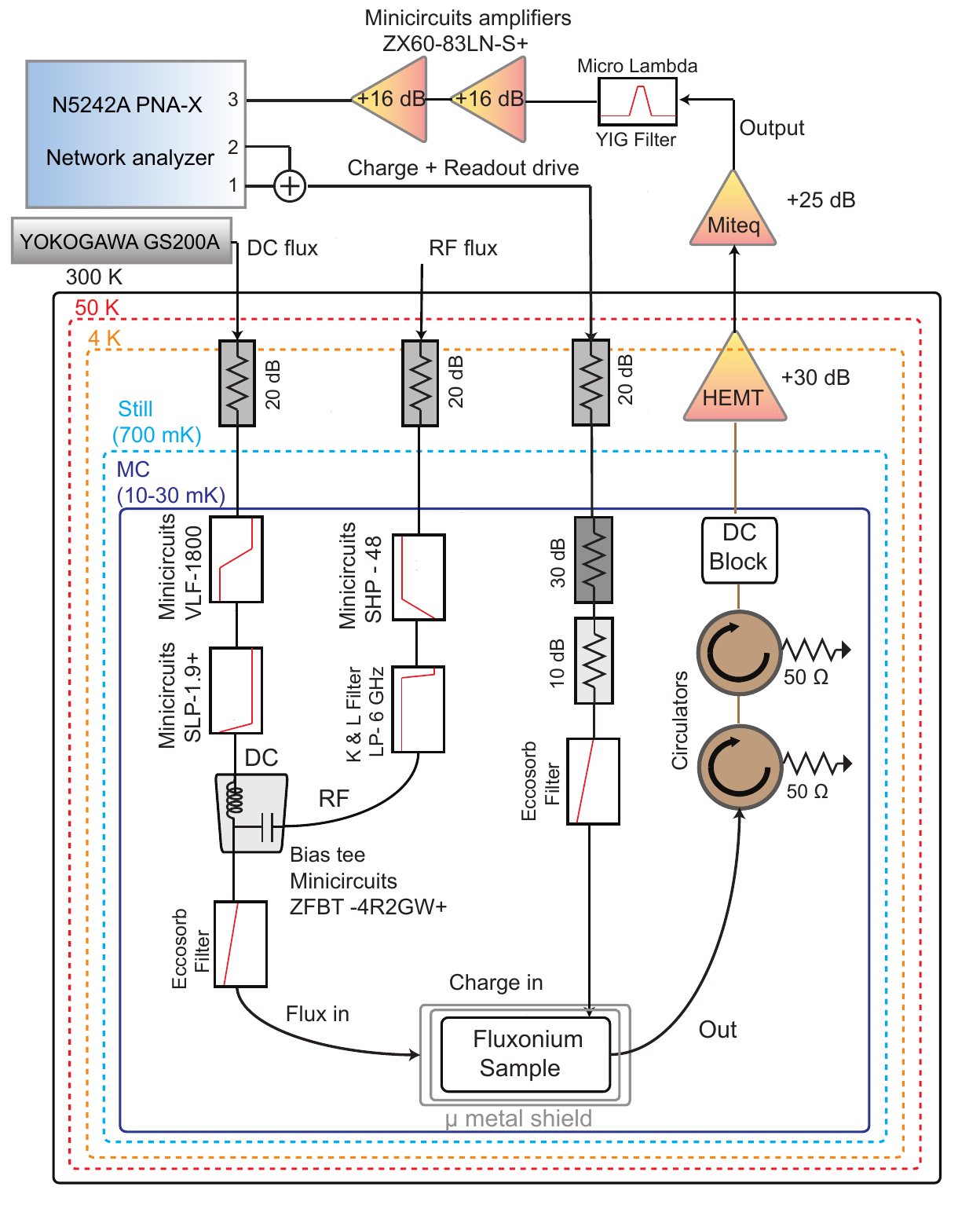}};

\end{tikzpicture}
  \caption{Wiring of microwave and DC connections to the device.
\label{wiring diagram}}
\end{figure}
The device (shown in Fig. 1 in the main text) was fabricated on a 430 $\mu$m thick C-plane sapphire substrate. The base layer of the device, which includes the majority of the circuit (excluding the Josephson junctions of the transmon), consists of 150 nm of niobium deposited via electron-beam evaporation at 0.9 nm/s, with features defined via optical lithography and reactive ion etch (RIE) at wafer-scale. The wafer was then diced into 7x7 mm chips. The junction mask was defined via electron-beam lithography with a bi-layer resist (MMA-PMMA) in the bridgeless junction pattern~\cite{lecocq2011junction}, with overlap pads for direct galvanic contact to the optically defined capacitors. Before deposition, the overlap regions on the pre-deposited capacitors were milled \textit{in-situ} with an argon ion mill to remove the native oxide. The junctions were then deposited with a three step electron-beam evaporation and oxidation process. First, an initial 30 nm layer of aluminum was deposited at 1 nm/s at an angle of -29$^{\circ}$ relative to the normal of the substrate. Next, the junctions were exposed to 50 mBar of high-purity $\text{O}_2$ for 45 minutes for the first layer to grow a native oxide. Finally, a second 60 nm layer of aluminum was deposited at 1 nm/s at 29$^{\circ}$ relative to the normal of the substrate. Another oxidation step at 3mbar for 5minutes was done after to put a clean oxide layer atop the aluminum. After evaporation, the remaining resist was removed via liftoff in N-Methyl-2-pyrrolidone (NMP) at 80$^{\circ}$C for 6 hours, leaving only the junctions directly connected to the base layer, as seen in the inset of Figure  1 of the main text. After both the evaporation and liftoff, the device was exposed to an ion-producing fan for 15 minutes, in order to avoid electrostatic discharge of the junctions. 

The device is mounted and wirebonded to a multilayer copper PCB microwave-launcher board. Additional wirebonds connect separated portions of the ground plane to eliminate spurious differential modes. The device chip rests in a pocketed OFHC copper fixture that presses the chip against the launcher board. Notably, the fixture contains an additional air pocket below the chip to alter 3D cavity modes resulting from the chip and enclosure, shifting their resonance frequencies well above the relevant band by reducing the effective dielectric constant of the cavity volume. The filtering, amplifier chain, and wiring diagram of microwave and DC and microwave connections to the device are as in Fig.~\ref{wiring diagram}.

\section{Equivalent Circuit}
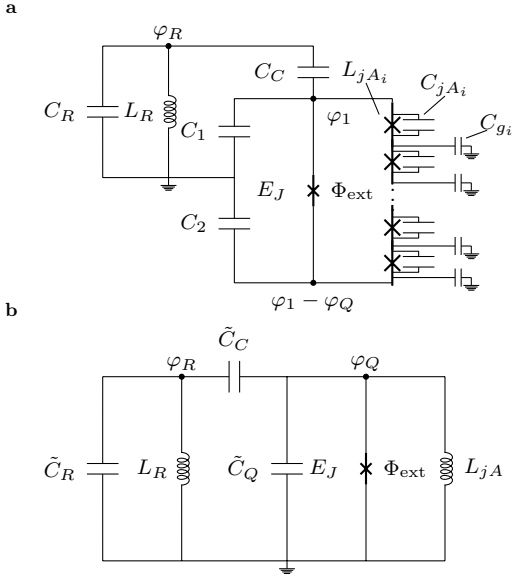
\begin{figure}[h!]
    \begin{tikzpicture}
    \node[anchor= north west] at (-7.0,0) {\usebox{\mycircuitb}};
    \node[scale=0.8] at (-7.2,0) {\textbf{a}};
    \node[anchor= north west] at (-7.0,-4.0) {\usebox{\mycircuitc}};
    \node[scale=0.8] at (-7.2,-4.0) {\textbf{b}};
    \end{tikzpicture}
    \caption{(a) The complete four node circuit model of the heavy fluxonium device shown in Fig.1. The inductive chain is made up of a series of large-area Josephson junctions with inductance $L_{jA_i}$ and capacitance $C_{jA_i}$, with each having a parasitic capacitance to ground, $C_{g_i}$. The idealized behavior of the inductive chain requires that each individual chain junction have a sufficiently large area to reduce both phase slips in the chain and the uncontrolled offset charges on the islands. Furthermore, the chain length must be short enough to avoid having the total stray capacitance $\sum_i C_{g_i}$shunting the chain inductance, but while also ensuring the total chain inductance, $L_{jA}$, is substantially larger than the individual junction's inductance, $L_J$. Requirements further detailed in ~\cite{manucharyan2009fluxonium}.  This circuit is equivalent to the three node circuit shown in (b), whose effective capacitances, $\tilde{C}$, are all defined in terms of those in (a), see Eqn.~\ref{Cres}, \ref{Cqubit}, and \ref{Ccoupling}.}
    \label{equivcirc}
\end{figure}
The actual circuit of the heavy fluxonium device is shown in 
Fig. \ref{equivcirc}(a). In this appendix, we 
exactly treat the realistic four-node circuit (assuming that we can replace
the Josephson chain as a pure inductor), and show that it reduces to an 
effective three-node circuit shown in the main text and again here 
[Fig \ref{equivcirc}(b)].  This circuit is described by the Hamiltonian 
$H_f$ (Eq. 2 of the main text), with effective three-node 
capacitances defined in terms of their four-node counterparts.

The phase variables for each of the circuits are shown at their respective nodes (small black circles). With one node as ground, the four-node ciruit has three phase variables, while the three-node has just two. 
The key point in this reduction is that the potential energy in both cases depends on only two variables, and we choose the labeling of nodes in Fig. \ref{equivcirc}(a) to make the two potential energies equivalent.  Thus, we need only show that the kinetic energy of \ref{equivcirc}(a) reduces in form 
to that of \ref{equivcirc}(b).  The kinetic 
energy $T$ is the capacitive energy, since the voltage at each node is $V_j = \hbar \dot{\varphi}_j/(2e)$.  For the four-node circuit, then, 
\begin{eqnarray}
 & T = \frac{1}{2} \left( \frac{\hbar}{2e} \right)^2 [C_R \dot{\varphi}_R^2 +
  C_C (\dot{\varphi}_1 - \dot{\varphi}_R)^2 \nonumber    \\
 & + C_1 \dot{\varphi}_1^2 + C_2 (\dot{\varphi}_1 - \dot{\varphi}_Q)^2 ]
\label{KE1}
\end{eqnarray}
The potential energy of the system, 
\begin{equation}
  V = \frac{1}{2} E_{LR} \varphi_R^2 + \frac{1}{2} E_{L{jA}} \varphi_Q^2
  + E_J (1 - \cos \varphi_Q),
\label{PE}
\end{equation}
is independent of $\varphi_1$, so that the Euler-Lagrange equation of motion,
$\frac{d}{dt}\frac{\partial L}{\partial \dot{\varphi}_1} = 
\frac{\partial L}{\partial \varphi_1} = 0$, where $L = T - V$ gives
\begin{eqnarray}
 & \partial L /\partial \dot{\varphi}_1 = \frac{\hbar}{2e} 
   [(C_1 + C_2 + C_C)] \dot{\varphi}_1  \nonumber   \\
  &  - C_C \dot{\varphi}_Q - C_2 \dot{\varphi}_Q ]
\label{constant of motion}
\end{eqnarray}
is a constant of the motion, which we may set to zero.  This allows us to
eliminate $\dot{\varphi}_1$ from Eqn.\ref{KE1}, which, after some
algebra, results in the desired form,
\begin{equation}
  T = \frac{1}{2} \left( \frac{\hbar}{2e} \right)^2 \left[\tilde{C}_R 
  \dot{\varphi}_R^2 + \tilde{C}_Q \dot{\varphi}_Q^2 -2 \tilde{C}_C
  \dot{\varphi}_R \dot{\varphi}_Q\right],  
\label{KE2}
\end{equation}
with three effective capacitances [corresponding to 
Fig. \ref{equivcirc}(b)] defined as
follows:  With the preliminary definition,
\begin{equation}
  C_T = C_1 = C_2 + C_C,
\label{Ctotal}
\end{equation}
the effective capacitances are 
\begin{eqnarray}
  \tilde{C}_R = C_R + C_C (C_1 + C_2) / C_T,  \label{Cres}   \\
  \tilde{C}_Q = C_2 (C_1 + C_C) / C_T,   \label{Cqubit}    \\
  \tilde{C}_C = C_2 C_C / C_T.          \label{Ccoupling}    
\end{eqnarray}

Physically, $\tilde{C}_Q$ is the capacitance between the two nodes of the
qubit, and $\tilde{C}_R$ is the capacitance between the resonator and ground nodes. 
We note that $\tilde{C}_C$ vanishes as $C_2 \rightarrow 0$ , and approaches $C_C$
as $C_2 \rightarrow \infty$.  In our system, $C_2 \approx C_1 + C_C$, so
$\tilde{C}_C \approx C_C/2$.

\section{Direct Drive}
\begin{figure}[h!]
    \begin{tikzpicture}
    \node[scale=0.8] at (-5.6,0) {\textbf{a}};
    \node[anchor=north west] (image) at (-6.6,0) {\includegraphics[scale=0.28]{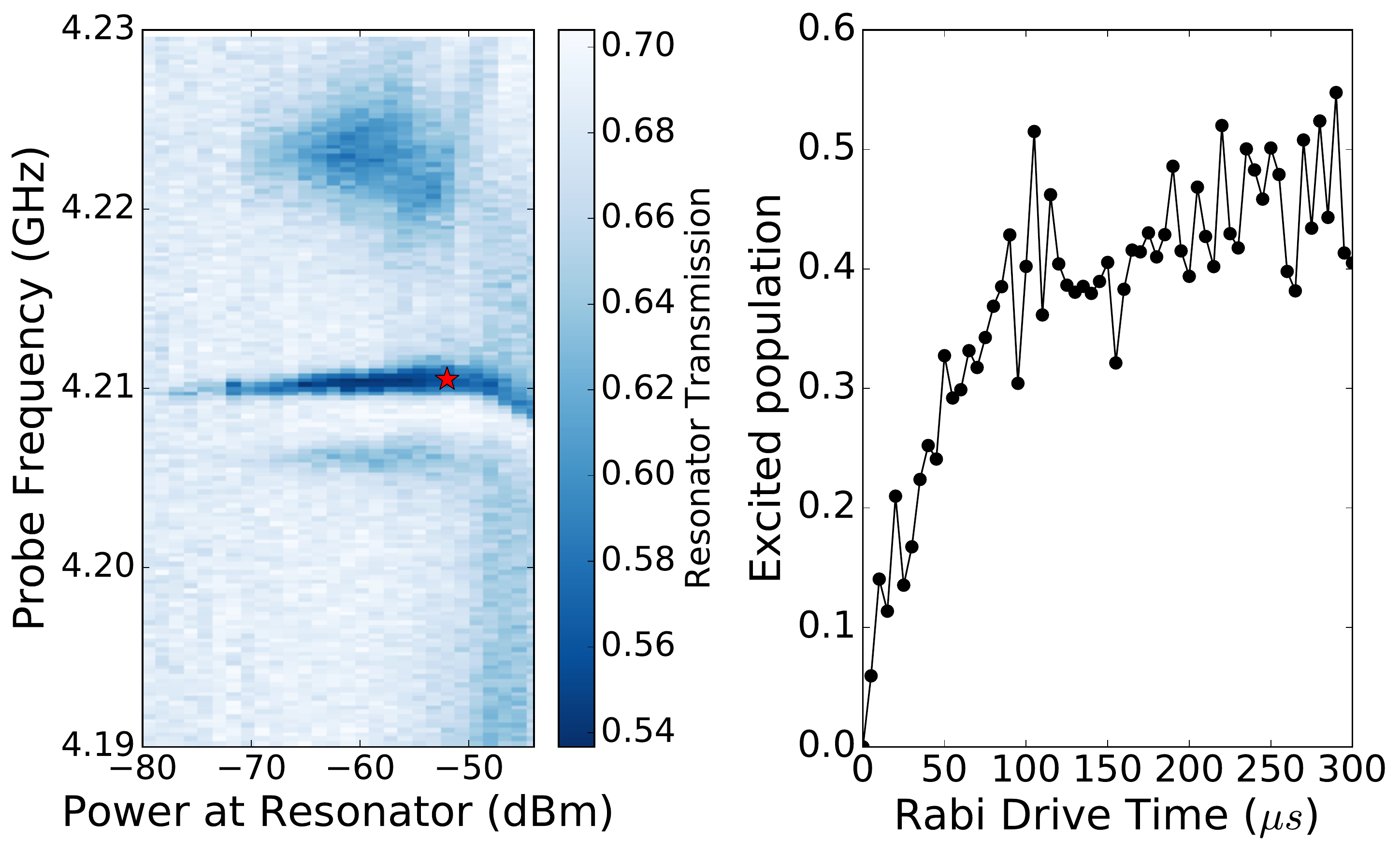}};
    \node[scale=0.8] at (-2.0,0) {\textbf{b}};       
    \end{tikzpicture}
    \caption{a) Power sweep of the probe tone while directly driving the fluxon transition at $\Phi = 0.078\Phi_{0}$. We draw attention to two features: 1) The width of fluxon transition ($4.21$\,GHz) saturates at high powers and 2) The appearance of other photon-assisted transitions immediately above and below the fluxon transition, pointing towards a source of limitation for the achievable Rabi rates.  b) At a power of -52 dB, where the fluxon transition has the largest width, a Rabi drive results in no coherent oscillations and a 50/50 mixture in $\sim$100 $\mu$s.    
\label{directpowersweep}}
\end{figure}

As mentioned in the main text, we cannot achieve coherent operations on the metastable states by driving directly on the fluxon transition. The result of such a drive at $\Phi = 0.078\Phi_{0}$ is shown in Fig.~\ref{directpowersweep}. The width of the fluxon transition is found to saturate at high drive powers, in Fig.\ref{directpowersweep}a.  Directly driving the fluxon at the location of largest width of the fluxon line results in the population saturating to a mixed state in $\sim 100$ $\mu$s. 

\section{Origin of Plasmon Dispersion}

The simulations of the spectra shown in Fig.~3 of the main text account accurately for the observed plasmon dispersion, and also for the smaller dispersion observed in the resonator
line.  In this and the next section we provide a simplified analysis to interpret and trace the origins of these and related phenomena.  This section is devoted to estimating the change in the bare qubit levels arising from $\Phi_{\rm ext}$ induced distortions of the potential well shape, with resonator interaction effects put off to the following section.

The strategy is to expand the potential about its shifted minimum to
characterize the distortion, and then apply perturbation theory to obtain the spectra. The resulting dispersion is quadratic in $\Phi_{\rm ext}$, with small quartic corrections.  

When $\Phi_{\rm ext}$ is applied, the central well rides up inside the inductive parabola, making contact at the phase value
$\varphi = 2\pi \Phi_{\rm ext}/\Phi_0$, where the cosine term vanishes.  
However, clearly, this point is not the potential minimum; the actual 
minimum is Stark-shifted by the inductive current, $E_L \varphi$, so that $\varphi_{\mathrm{min}} = 2\pi (\Phi_{\rm ext}/\Phi_0) (1-E_L/E_J)$, to lowest 
order in $E_L/E_J$.  We define the displacement from this minimum as $\tilde{\varphi} = \varphi - \varphi_{\mathrm{min}}$, and rewrite the potential 
energy as a function of $\tilde{\varphi}$:
\begin{equation}
  V = E_J \big[1- \cos \big(\tilde{\varphi} - 2\pi {E_L \over 
  E_J}{\Phi_{\rm ext} \over  \Phi_0} \big) \big] 
  +  {1 \over 2} E_L (\tilde{\varphi} + \varphi_{\mathrm{min}}) ^2.
\label{new potential 1}
\end{equation} 
Expanding and ignoring constant terms,
\begin{eqnarray}
 & V = -E_J \big( \cos \tilde{\varphi} \cos 2\pi {E_L \over E_J}{\Phi_{\rm ext}  \over  \Phi_0} + \sin \tilde{\varphi} \sin 2\pi 
  {E_L \over E_J}{\Phi_{\rm ext} \over  \Phi_0} \big)   \nonumber   \\ 
 & + {1 \over 2} E_L \tilde{\varphi}^2 + E_L \varphi_{\rm min} \tilde{\varphi}.
\label{new potential 2}
\end{eqnarray}
Making small-angle expansions and regrouping,
\begin{eqnarray}
 & V = -E_J \big[ 1 - 2 \big( {\pi E_L \over E_J} \big)^2}
  \big( {\Phi_{\rm ext} \over \Phi_0} \big)^2 \big]~\cos {\tilde{\varphi} \nonumber
  \\  & + {1 \over 2} E_L \tilde{\varphi}^2 + 2 \pi E_L 
 \big( {\Phi_{\rm ext} \over \Phi_0} \big)~(\tilde{\varphi} - 
 \sin \tilde{\varphi}).
\label{new potential 3}
\end{eqnarray}
This equation shows that the symmetric part of the displaced well is identical to the original well, except for a reduction in the effective $E_J$ by the small relative amount shown, while the odd part appears and cancels the
linear part of the inductive term, leaving cubic and higher odd powers.

Both symmetric and antisymmetric parts of $V$ contribute quadratic terms
($\sim \Phi_{\rm ext}^2$) to the plasmon dispersion; the even part at first order in perturbation theory, the odd part at second order.  To estimate the former, note that in the harmonic approximation, the plasmon energy $E_P$ increases 
as $\sqrt{E_J}$, so that the fractional reduction is half that of $E_J$, and
hence
\begin{eqnarray}
  \Delta E_P (\hbox{symm}) &= - \big({\pi E_L \over E_J} \big)^2 E_P
  \big(\Phi_{\rm ext}/\Phi_0 \big)^2   \nonumber   \\   
 & \approx -45 \hbox{MHz}~\big(\Phi_{\rm ext}/\Phi_0 \big)^2.
\label{shift even}
\end{eqnarray}

Regarding the antisymmetric contribution, applying second-order 
perturbation theory to the leading cubic term in $\tilde{\varphi}$ gives:
\begin{eqnarray}
 \Delta E_P (\mathrm{anti}) &=& \big(\frac{\pi}{3} E_L \big)^2 
 \left( {\Phi_{\rm ext} \over \Phi_0} \right)^2 \cdot   \nonumber   \\
 &&  \left( { |\langle f_0 | \varphi^3 | e_0 \rangle |^2 \over E'_P } 
 - 2 \frac{|\langle g_0 | \varphi^3 | e_0 \rangle|^2}{E_P} \right)\nonumber
\label{second order}
\end{eqnarray}
The first term in brackets is the shift in energy of the $\ket{e_0}$ state due
to virtual occupation of $\ket{f_0}$, with energy denominator $E'_P = E_{f_0} - E_{e_0} = 4.39$\,GHz.  Virtual occupation of $\ket{g_0}$ accounts for half the second term.  The other half is (minus) the shift in energy of $\ket{g_0}$ due to virtual occupation of $\ket{e_0}$, with energy denominator $E_P = 5.072$\,GHz.  Evaluating 
the matrix elements using $\varphi = 0.60 (a + a^{\dagger})$, we find
\begin{equation}
  \Delta E_P (\hbox{anti}) \approx 39 \hbox{MHz}~
  \big(\Phi_{\rm ext}/\Phi_0 \big)^2.
\label{shift odd}
\end{equation}
The above two contributions together predict a total quadratic dispersion 
$\Delta E_P = 84$~MHz~$(\Phi_{\rm ext}/\Phi_0)^2$.  Quantitative agreement with the observed dispersion is deceptive - interaction with the resonator reduces the prediction by about 12 \%.  While quantitative agreement is not expected or sought from such a calculation, it suggests that a predominantly quadratic plasmon dispersion arises from $\Phi_{\rm ext}$ induced distortions in the well, with
comparable contributions arising from reduced effective $E_J$ in the 
symmetric part, and induced antisymmetry dominated by cubic anharmonicity. Quartic corrections arise at the next contributing order in perturbation 
theory, second order for the symmetric part, and fourth order for the
antisymmetric part. We have found these to be no greater than 1\% of the quadratic contributions. [The small parameter governing successively higher contributing orders in perturbation theory in both cases is $(E_L/E_P)^2$.]  Larger (but still small) quartic corrections will arise from the interaction with the resonator, as shown in the following.

\section{Resonator/Qubit hybridization}

The most direct indication of the strong interaction between the principle plasmon and the resonator is provided by the power sweep:  At low power levels one sees both features, separated by 155\,MHz, whereas at higher powers, one sees only the resonator line, shifted upward by 33\,MHz.
\begin{figure}
\includegraphics[scale=0.3]{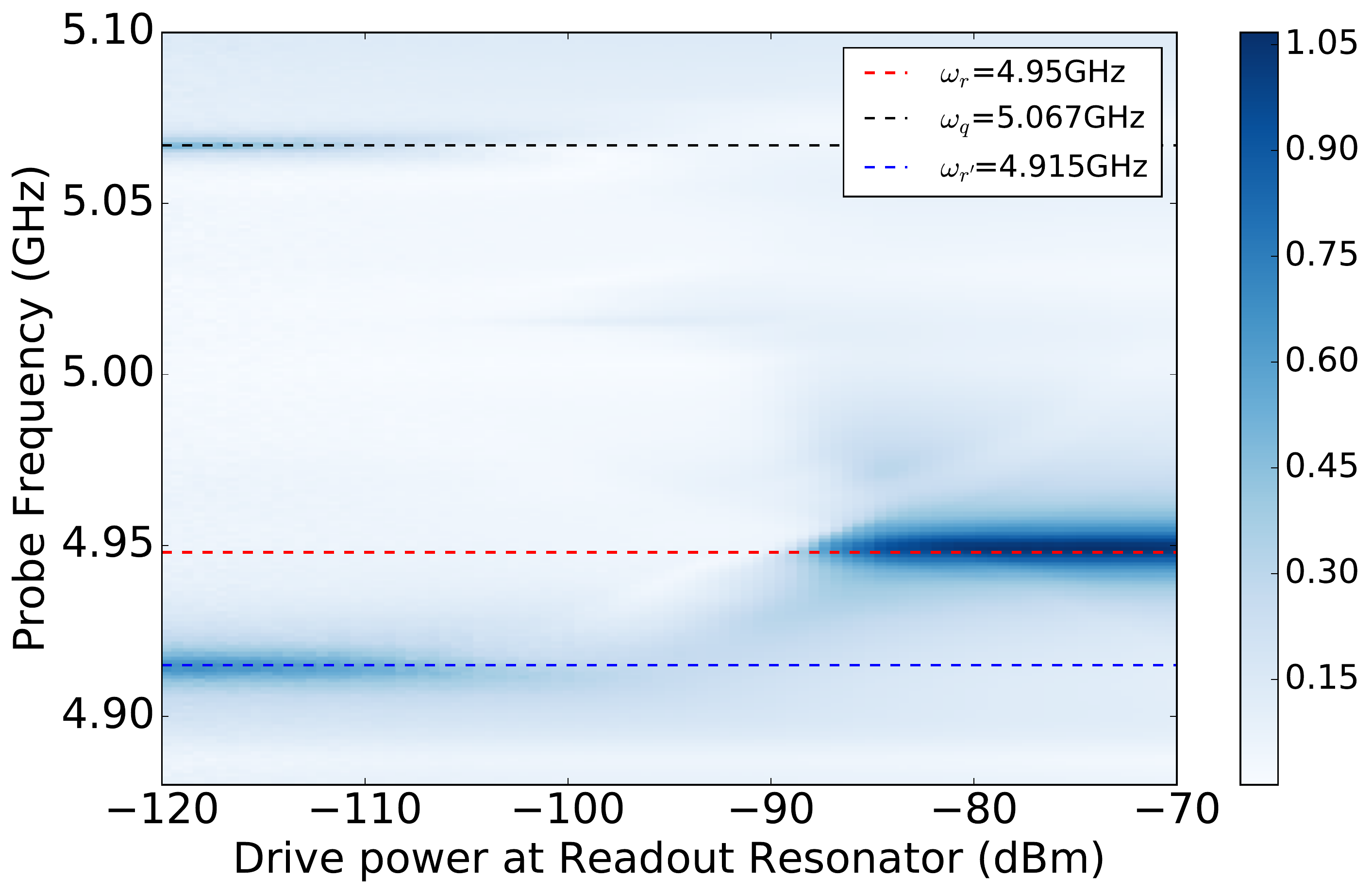}
\caption{Single Tone Spectroscopy power sweep at the input of the readout resonator. At high powers ($> -90$\,dB), the heavy fluxonium qubit is saturated, and the bare resonance, $\omega_r$ of the readout resonator is seen. At lower powers ($< -100$\,dB) the qubit plasmon $\omega_q$ and the dressed resonator, $\omega_{r'}$ are seen. The bare resonator transmission amplitude is used to normalize the spectra in all the figures of the main text.
\label{powersweep}}
\end{figure}
Assuming that the high-power value is the upcoupled (bare) resonator line, one can infer the interaction strength $g$ from these three numbers alone. This is found to be within the range of experimental uncertainty of the value $g = 76$\,MHz  (Fig.~\ref{powersweep}) determined by a global fit to all the data.

Here we introduce a simple model that describes the plasmon and resonator lines over the entire flux range, excluding the immediate neighborhood of the points $\Phi_{\rm ext} = 0$ and $\Phi_0/2$, where more transitions are involved.  Consider the coherent superposition of the two lowest excited states of the noninteracting system, one involving the resonator and the other involving the qubit:
\begin{equation}
   |\psi\rangle = \alpha |1_R,g_0\rangle + \beta |0_R,e_0\rangle.
\label{twostate}
\end{equation}
The effective Hamiltonian acting on the subspace of \ref{twostate} is
\begin{equation}
H = \left(\begin{array}{cc}
\epsilon_R & m   \\
m^{*} & \epsilon_{Q} \end{array}\right)
\label{twomatrix}
\end{equation}
where $m$ is the interaction matrix element specific to the plasmon excitation, $m = g \langle e_0 |n| g_0 \rangle = 0.062 i$. 
Diagonalizing $H$, we find the two lowest excited eigenstates of the system.  Near $\Phi_{\rm ext} = 0$, where the bare detuning is $\epsilon_Q - \epsilon_R \approx 89$\,MHz, we find the resonator-like eigenstate defined by $(\alpha/\beta)_R =
1.94i$, and the qubit-like eigenstate $(\alpha/\beta)_Q = -i/1.94$.  In the 
former, there is a probability of $|\beta_R|^2 = 0.21$ that the qubit will be found in the $(\ket{e_0}$ state, and conversely, in the latter, the same probability that the resonator will be found in its $\ket{1}$ state.  

Since both resonance lines are read from the resonator, their relative intensities are given by the relative probabilities of finding the resonator in its $\ket{1}$ state,
\begin{equation}
   I_R/I_Q = |\alpha_R/\alpha_Q|^2 = 3.8,
\label{relative intensities}
\end{equation}
and a rough comparison of resonance peak heights times linewidths is
consistent with this estimate. 
As $\Phi_{\rm ext}$ is increased and the 
detuning is reduced, the eigenstates exhibit further hybridization.  At
$\Phi_{\rm ext} = 1/2$, for example, the predicted intensity ratio is reduced
to $I_R/I_Q = 2.6$.

A more obvious consequence of hybridization, quantitatively, is the 
curvature of the resonator line, which appears to be due entirely to
level repulsion from the qubit.  To demonstrate, we write the bare
plasmon energy as $\epsilon_Q = 5.039$ GHz $- a (\Phi_{\rm ext}/\Phi_0)^2$,
where a purely quadratic dependence on $\Phi_{\rm ext}$ is assumed, following arguments of the previous section.  Assuming that $\epsilon_R$ is a constant, we may write down the eigenvalues of $H$ and evaluate the leading quadratic dependence on $\Phi_{{\rm ext}}$ with small quartic corrections.  The
$\Phi_{\rm ext}$ dependent parts are
\begin{eqnarray}
  & \Delta E_R = - 0.21 a~[\Phi_{\rm ext}/\Phi_0]^2
   - 0.3 a~[\Phi_{{\rm ext}}/\Phi_0]^4,  
\label{dispersionR}  \\
  & \Delta E_Q = - 0.79 a~[\Phi_{\rm ext}/\Phi_0]^2
   + 0.3 a~[\Phi_{{\rm ext}}/\Phi_0]^4.  
\label{dispersionQ}
\end{eqnarray} 
While the estimate of the previous section may not be accurate quantitatively,
the ratio of coefficients found here should be (of both the quadratic and
quartic terms).  It is interesting to note that the ratio of quadratic 
coefficients is 3.8; compare that of \ref{relative intensities}.  

The calculation above supports the interpretation that the observed dispersion
of the resonator line is a direct consequence of hybridization with the plasmon, whose dispersion is driven in turn by $\Phi_{\rm ext}$ induced distortions in the potential well.  And, like the plasmon, the resonator line has two distinct branches visible near the half flux quantum point, one generated by hybridization with the $\ket{g_0} \rightarrow \ket{e_0}$ branch, the other with the $\ket{g_1} \rightarrow \ket{e_1}$ branch. 

\section{Photon/Plasmon Assisted Transitions}

\begin{figure}
\begin{tikzpicture}
\node[anchor=north west] (image) at (-1.5,0){\includegraphics[scale=0.28]{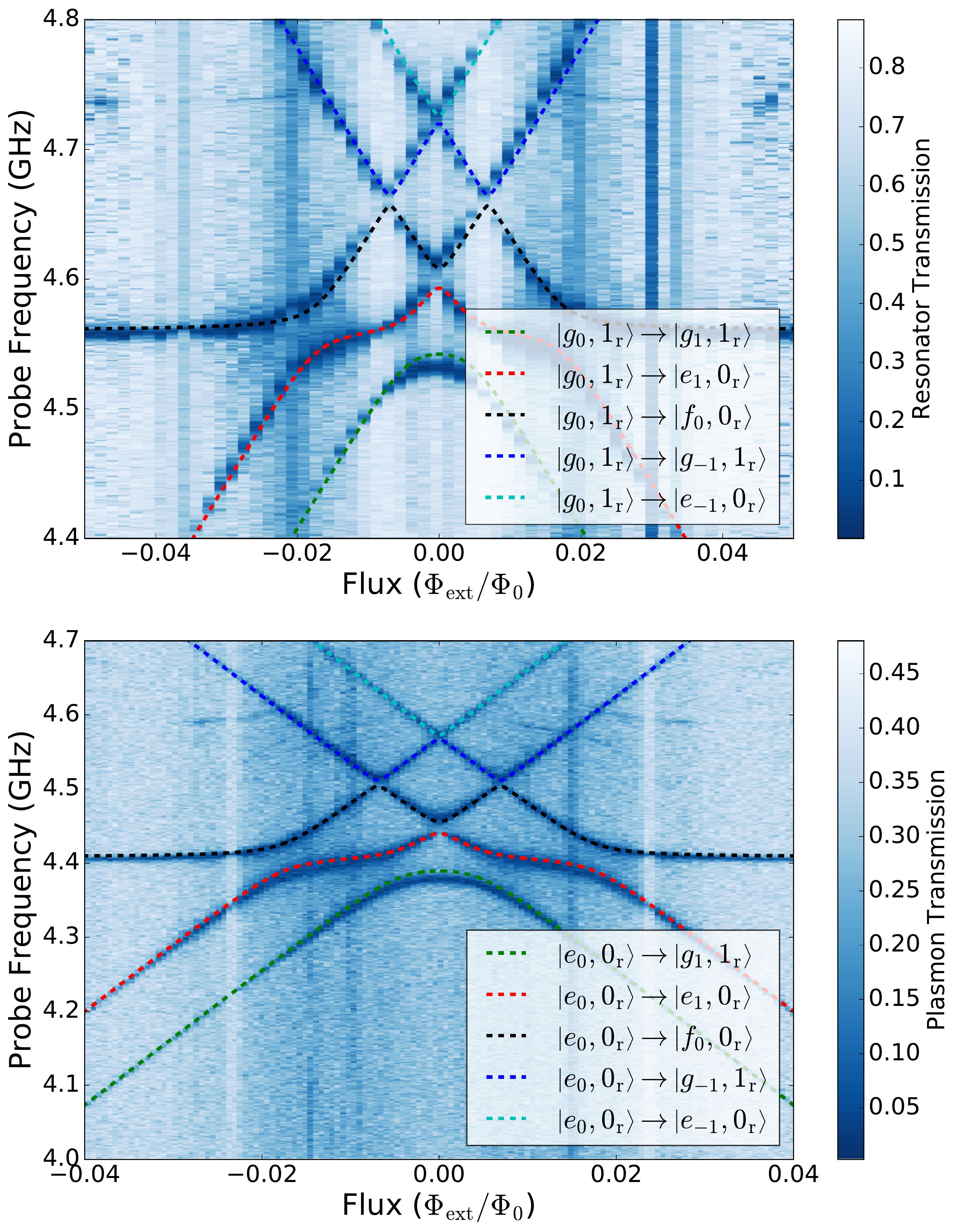}};
\node[scale=0.8] at (-0.25,0) {\textbf{a}};
\node[scale=0.8] at (-0.25,-5.25) {\textbf{b}};
\end{tikzpicture}
\caption{ Two-tone spectroscopy as a function of flux (near zero flux) showing (a) resonator-photon and (b) plasmon assisted transitions. In (a) the readout tone is placed at the resonator frequency and the transitions start with the fluxonium-resonator system in $\ket{g_{0}}\otimes\ket{1}$. In (b), the `readout' tone is placed at the plasmon frequency and the transitions start with the fluxonium-resonator system in $\ket{e_{0}}\otimes\ket{0}$.  In each case, the transmission of a readout tone is monitored, while the frequency of a second drive tone is swept (y-axis). Both sets of transitions are absent when the drive and readout tones are pulsed and staggered, as a result of the short lifetimes of both the resonator and the plasmon. (State labeling for the transitions is valid for $\Phi_{\text{ext}} > 0$.)}
\label{photonplasmonassist}
\end{figure}

The two-photon features of Fig.~4a  of the main text are a result of transitions to the two-excitation subspace ($\ket{e_{1,-1}},\ket{f_{0}}$) starting with the combined fluxonium-resonator in $\ket{g_0}\otimes\ket{0}$ state. The avoided crossing associated with this two-excitation space are also observed in the resonator-photon and plasmon assisted transitions shown in Fig.~\ref{photonplasmonassist}a, b. These transitions start with the coupled resonator-fluxonium system in $\ket{g_0,1_{\text{res}}}$ and $\ket{e0,0_{\text{res}}}$, respectively. The final states of the transition are the same as those in Fig.~4a of the main text, but are here reached using a single photon, with the second photon acquired from the readout resonator and the plasmon, respectively. Due to the low Q  of the readout resonator (Q $\sim$ 500), and the resulting Purcell limited lifetimes of the primary plasmon features, these transitions cannot be used for coherent transitions between the states. Increasing the Q of the readout resonator, and it's detuning from the primary plasmon will allow the use of these transitions for coherent operations between fluxoid states, as will be explored in future devices. 

The resonator assisted transitions are useful for photon `latching' in which a single photon from the resonator can be transferred to an excited metastable state, and subsequently detected by cycling on the excited plasmon transition. This might be potentially useful in dark photon detection schemes. 

\section{Qubit Inversion in Single Tone Spectroscopy}
\begin{figure}
\includegraphics[scale=0.26]{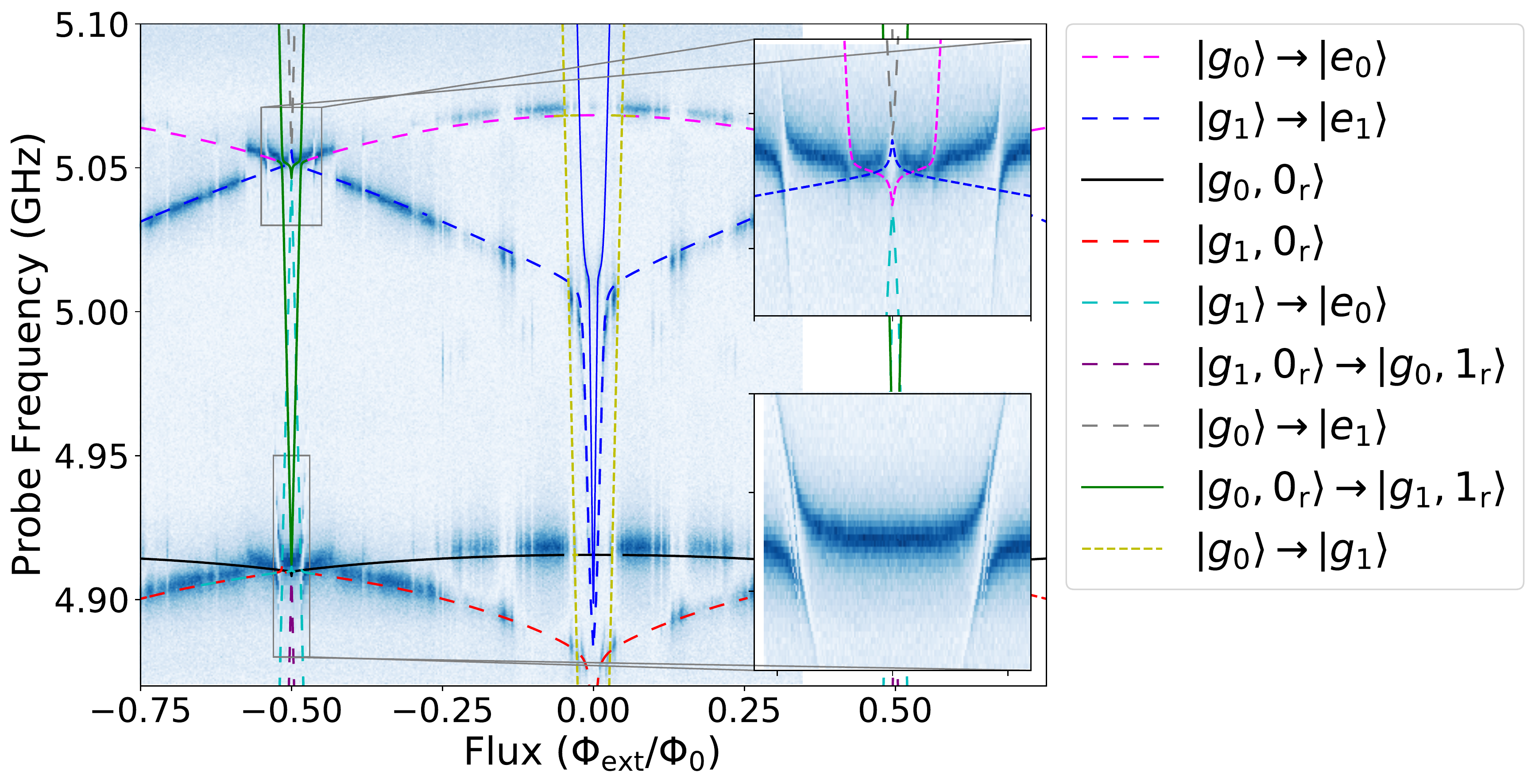}
\caption{Single tone spectroscopy demonstrating qubit state inversion for varying time durations, and appearance of new avoided crossings. Reasons for this inversion and avoided crossings are unknown and will be part of future investigations.}
\label{inverted}
\end{figure}

During the measurement of the device, there was a week long period in  which the device was in a frustrated excited state as shown Fig.\ \ref{inverted}. We take particular note of the switching from the $\ket{g_1} \rightarrow \ket{e_1}$ plasmon branch to the $\ket{g_0} \rightarrow \ket{e_0}$ branch. While we were unable to reproduce this particular spectrum, with inverted single tone spectrum and the onset of new avoided crossings, the switching from one well branch to another seems to have been a common feature of this device. This switching can also been seen at $\Phi_{\text{ext}}\approx 0.6\Phi_0$ in Fig. 2 in the main text. A more careful analysis to address the exact nature of the metastable qubit inversion is warranted.  Fig.~\ref{inverted}, however, is a further demonstration of the large dispersive shifts of the readout resonator, and demonstrates the ability to readout the fluxoid state over the entire flux range. 

\section{Steady-State Simulations for Single-Tone Spectroscopy}
We perform numerical simulations of single-tone spectroscopy using the Lindblad master equation:
\begin{align}\label{masterEquation}
   \frac{d\rho (t)}{dt} =& -i \left [H(t),\rho (t) \right]  \\\nonumber
   &+\sum_{\omega} \gamma_{\omega} (T) \left( A_\omega  \rho(t) A_{\omega}^\dagger - \frac{1}{2} \{   A_{\omega}^\dagger A_\omega,\rho(t) \}  \right).
\end{align}
Here, $\gamma_{\omega}$ is the temperature-dependent decoherence rate for a transition with frequency $\omega$, $A_\omega$ is the corresponding jump operator, $\rho$ the density matrix, and $H(t)$ the Hamiltonian describing the coupled fluxonium--resonator system along with a coherent drive: 
\begin{align}\label{drivenHam}
H(t) = &H_{\mathrm{Fluxonium}} + H_{\mathrm{Resonator}} \\\nonumber 
&+\sum_{n,m} g_{nm} \ket{n}\bra{m} \left(a + a^\dagger \right) + \zeta \left( a e^{i \omega_d t} + \mathrm{h.c.} \right)
\end{align}

The drive strength is small, i.e., $\zeta \ll \omega_d$, and the rotating wave approximation between the driving field and resonator is used to drop rapidly oscillating terms. The qubit coupling strengths, $g_{nm}$, are obtained from the fluxonium charge matrix elements. The dissipation operators, $A_\omega$, are eigenoperators of the qubit-resonator system (not including the driving field). The temperature-dependent decoherence rates, $\gamma_\omega (T)$, are given by a resonator rate, $\kappa f(T,\omega)$, and fluxonium rates, $\Gamma |\bra{n}N\ket{m}|^{2} f(T,\omega)$. $\bra{n}N\ket{m}$ are fluxonium charge matrix elements, $f(T,\omega > 0) = e^{\beta\omega}/(e^{\beta\omega}-1)$, and $f(T,\omega < 0) = 1/(e^{\beta\omega}-1)$ . $\Gamma$ and $\kappa$ are rate constants that are fit from the experimental transmission data near $\Phi_{\text{ext}} = 0$. These rate constants are found to be $\Gamma = 0.0005$ GHz and $\kappa = 0.04$ GHz.  

\begin{figure}
\includegraphics[scale=0.8]{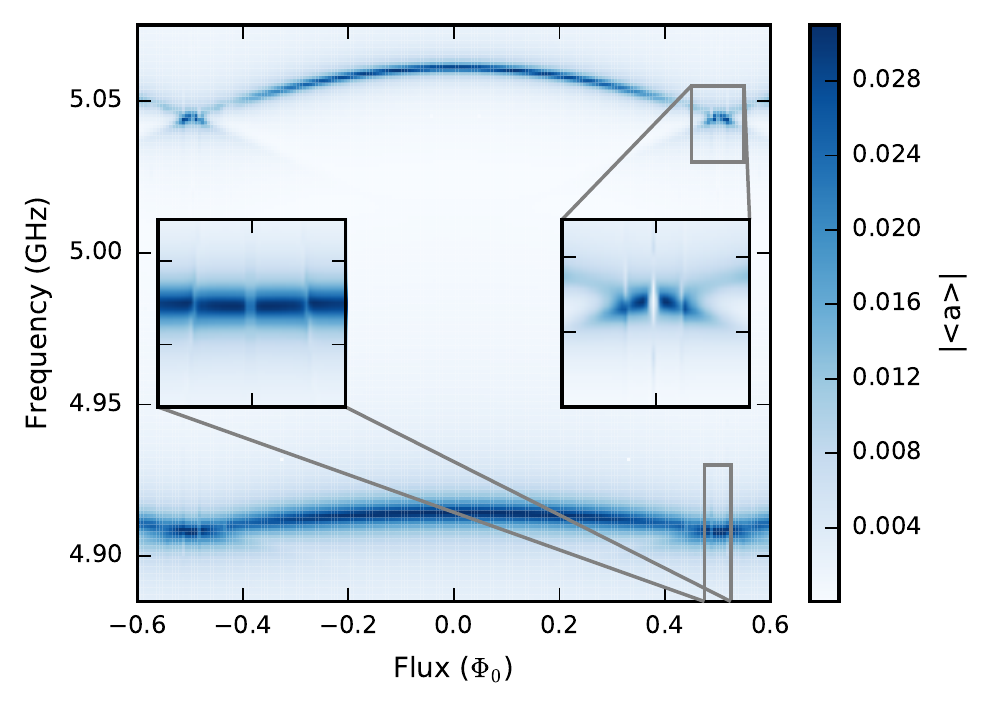}
\caption{Steady-state single tone simulation of the coupled fluxonium-resonator system at a finite temperature of 30mK. When a non-zero temperature is included in the steady-state simulation, a non-zero photon expectation value in the $\ket{g_1} \rightarrow \ket{e_1}$ branch is observed, agreeing with the single-tone measurements shown in Fig.\ 2.}.
\label{steadystate}
\end{figure}

The signal transmission amplitude is obtained  by calculating the average of the resonator annihilation operator, $|$tr(a$\rho$)$|$, using the solution to $\ref{masterEquation}$ in the long-time limit: $\frac{d\rho_{ss}'(t)}{dt} = 0$. We perform the calculation in a rotating frame in which time dependence of $\ref{drivenHam}$ is effectively eliminated. This removes the need for time averaging, considerably speeds up computation time, and thereby allows us to investigate behavior on time scales of milliseconds and greater, such as the the thermally assisted $\ket{g_1} \rightarrow \ket{e_1}$ transition depicted in figure $\ref{steadystate}$. 

\end{document}